# Mac Users Do It Differently: the Role of Operating System and Individual Differences in File Management

**Jesse David Dinneen**
Victoria University of Wellington
Wellington, New Zealand
jesse.dinneen@vuw.ac.nz

**Ilja Frissen**
McGill University
Montreal, Canada
ilja.frissen@mcgill.ca

## Abstract
Despite much discussion in HCI research about how individual differences likely determine computer users' personal information management (PIM) practices, the extent of the influence of several important factors remains unclear, including users' personalities, spatial abilities, and the different software used to manage their collections. We therefore analyse data from prior CHI work to explore (1) associations of people's file collections with personality and spatial ability, and (2) differences between collections managed with different operating systems and file managers. We find no notable associations between users' attributes and their collections, and minimal predictive power, but do find considerable and surprising differences across operating systems. We discuss these findings and how they can inform future research.

## Author Keywords
Personal information management; Computer files; Personality traits; Spatial ability

## CCS Concepts
•**Human-centered computing** → **Empirical studies in HCI;** *Graphical user interfaces;* •**Social and professional topics** → **User characteristics;** •**Applied computing** → **Document management;**

## Introduction

Contemporary computer use – in traditional, mobile, and embedded contexts – often requires users to invest considerable time and effort in managing computer files [8, 17]. Supporting this activity (e.g., through improved software) requires first understanding it, and so many prior works have examined users' perceptions of and challenges encountered while managing files, the file collections produced, and various other personal, social, and technological influences [8]. In particular, some studies have considered individual differences among users such as personality traits [2, 20], cognitive and spatial abilities [1, 22], and psychological traits [13, 19], but also technological determinants such as the software used to manage files [3]. Together, such studies are inconclusive, leaving notable gaps in our knowledge about the role of personality, spatial ability, and software in shaping people's file management behaviour and the collections that behaviour produces. This manuscript therefore extends a 2019 CHI paper [9] by providing additional analyses of existing data to explore the roles of these determinants in how people manage computer files.

## Related works

Personal information management (PIM) is the study and activity of (primarily) keeping, organising, and finding or re-finding personal information [17]. Among the most common forms of PIM is file management (FM), where users *keep* information by storing files, *organise* it by arranging files and folders, and re-find it by navigating or searching through the folder structure they have created [8]. Many aspects of FM have been examined in prior works, including users' perceptions and attributes, the tools they use to manage files, and the artefact (i.e., file collection) produced by their actions [8]. In particular, as PIM seems deeply personal and psychological [19], many works have reasoned there is an impact of, and called for further study of, the role of individual differences – what makes two people different with regards to their personality, cognition, and so on – in determining PIM and FM behaviour [8, 13]. Empirical validation of these ideas has been confined to just a few studies, however, which have primarily focused on peoples' personalities and spatial abilities [8].

Personality is typically conceptualised as grouped traits (most commonly, the Big 5: openness, conscientiousness, extraversion, agreeableness, and neuroticism [12]) or individual traits like a need for control, resistance to stress, and so on. One work [20] examined properties of file collections that participants assumed would predict Big 5 traits, with mixed results: though neuroticism and openness were not associated with collections' properties, conscientiousness was negatively associated with the number of files kept (i.e., conscientious participants kept fewer files), positively with the average number of files per folder (i.e., fewer folders with more files per folder), and positively with files on the desktop. However, associations were only under constrained analyses and among collections managed with particular operating systems, suggesting a role of software in determining participants' collections. Another study [2] examined the potential role of the Big 5 traits and several others (e.g., memory, computer literacy, orderliness) in determining how people *retrieve* their computer files, and found no significant associations. Taken together, these results are inconclusive, suggesting a need for further study of both personality and the role of software in determining FM behaviour and collections.

Spatial ability is another promising candidate for understanding what influences FM; collections are presented in two-dimensional space through which users *navigate*, causing activation in the areas of the brain responsible for spatial cognition [1]. One study found that participants with low spatial ability took twice as long to complete folder navigation tasks with command-line interaction [22], though some navigation issues could be partly alleviated by visualising the local position within the folder tree [23]. Since the command-line paradigm for file interaction has been largely supplanted by graphical file managers, it remains unclear if (and to what extent) spatial ability influences how people create and navigate folder structures today.

Finally, the role of software in influencing PIM and FM has been suggested by several studies, particularly because the tools (e.g., file managers) and software environments (e.g., operating systems or OS) adopted during some task enable, restrict, and affect user behaviour [4, 11, 18]. One study [3] explicitly examined the role of OS in FM, and found that participants retrieved files from their Mac computers faster than participants did on their own Windows computers. This difference appeared to be the result of differing organisational strategies – Mac users kept more folders closer to the root of the folder tree, with fewer files but more subfolders per folder – again suggesting it would be useful to take a deeper look at the differences between collections managed with different operating systems.

*Problem summary and research questions*
In summary, people's personalities, spatial abilities, and the software they use to manage files are promising, often-discussed factors for understanding and thus eventually better supporting FM, but the extent of their influences on FM, especially on users' collections, remains unclear. The research questions for the present study are therefore:

RQ1  Which (if any) measures of file collections' scale and structure are associated with or predicted by personality traits (e.g., the Big 5)?

RQ2  Which (if any) measures of file collections' scale and structure are associated with or predicted by spatial ability (e.g., sense of direction)?

RQ3  What (if any) are there notable and significant differences among the file collections managed with different operating systems or file managers?

**Methodology**
Quantitative descriptions of 348 participants' local file collections, presented in prior work [9], were analysed together with participants' responses to personality and spatial ability questionnaires.

*Data collection and preparation*
Remote and anonymous participants downloaded and ran data collection software described and validated in prior work [10, 9, 7], which presented questionnaires and scanned locations on participants' computers where they indicated they managed files. Care was taken to exclude files not managed by the participant: operating system folders were blacklisted (e.g., *Win32* and the Mac *Library* folder) and hidden files and folders were ignored. A summary of the participants and their computers is in Table 1, and the sample's heterogeneous makeup is discussed in [9]).

Measures of collections (defined and contextualised in [9]) used in the analysis included three measures of scale (total files, total folders, size in GB), nine measures of folder structure (# of roots and of root level folders, tree height, tree breadth, mean folder depth, branching factor, tree waist depth, % of leaf folders, and % of switch folders), and five measures of file structure/organisation (# of unfiled files, mean files per folder, mean file depth, % of empty folders, and depth of the file waist). The Big 5 personality traits were measured using the Ten-Item Personality Index [12], and spatial ability was measured using the Santa Barbara Sense of Direction Scale [14].

|  | male | female | other |
|---|---|---|---|
| gender | 218 (63%) | 123 (35%) | 7 (2%) |
|  | range | mean | SD |
| age | 14-64 | 30 | 9.9 |
|  | MacOS (10.8-.13) | Windows (XP-10) | Linux (8 distros) |
| OS | 169 (48%) | 135 (39%) | 44 (13%) |
|  | laptop | desktop | other |
| form | 263 (75%) | 82 (24%) | 3 (1%) |
|  | mixed | work/school | personal |
| use | 254 (73%) | 55 (16%) | 39 (11%) |

**Table 1:** Summary of the demographic (top) and technological (bottom) features of the population sample ($n$=348) and data set (approx. 50 million files and 8 million folders).

*Data analysis*
To analyse associations of file collection measures with personality traits and sense of direction, correlation matrices were produced. To ensure results comparable with prior works [20], one matrix was produced by log-transforming skewed data and performing parametric correlation tests (Pearson's rho) on all variable pairs. However, in light of recent discussions about analysing skewed data and the problems of log-transforming zero values [9, 7], two additional matrices were produced, one with and one without outliers (identified with interquartile range [24]), both with a non-parametric correlation test (Kendall's tau), which performs well with skewed, continuous data and does not require log transformation [6]. To explore the predictive power among variables, multiple regression models were generated with the personality dimensions and sense of direction as independent variables and the FM measures as dependent variables, log transforming and removing outliers as appropriate for each variable's distribution.

To analyse differences in file collections' scale and structure across operating systems, we performed Kruskal-Wallis tests across the three OS groups for each variable to identify any significant ($p$<0.05) difference among all three, then for each variable with a difference we ran pairwise parametric ($t$) or non-parametric (Mann-Whitney U) tests of difference (depending on the distribution) for each OS pair [7]. Outliers were removed when appropriate.

## Results

Across the correlation matrices generated to examine associations between file collection scale and structure with personality traits and with spatial ability (i.e., 96 associations analysed three different ways, as described above), we observed only correlation strengths between 0.16 and -0.18 ($p$<0.005 at those bounds), with most approaching zero (i.e., only very weak correlations). The multiple regressions showed that sense of direction, conscientiousness, and agreeableness could together somewhat predict a few measures of FM (most notably: total files, total folders, and folder tree breadth) with statistical significance (variously, $p$<0.05 and $p$<0.01), but the effect sizes were small ($R^2$<0.074, i.e., the predictors account for less than 7.4% of variation in the data) and no other combinations of variables were noteworthy. The full correlation and regression model results are shared online[1].

Across collections managed with the three examined operating systems we observed no meaningful differences (regardless of significance) among the number of roots (1-2) nor for percentage of leaf folders (72-73% for each OS), but almost every other measure of scale and structure shows statistically significant and often large differences. Table 2 shows notable differences, while the full tabular data are available online[1], and a summary follows.

---
[1] https://github.com/jddinneen/fm-results-tables

|  | Mac |  | Win. |  | Linux |  |
|---|---|---|---|---|---|---|
| total files | 106000 | *** | 45000 |  | 61000 | * |
| total folders | 19000 | *** | 4000 |  | 6000 | *** |
| size in GB | 109 | *** | 59 | * | 117 |  |
| # root folders | 15.5 | *** | 22 | * | 14.7 |  |
| branching factor | 5 | * | 3.8 |  | 4 |  |
| tree breadth | 5100 | *** | 1000 |  | 1400 | *** |
| tree depth | 17 | *** | 14 |  | 14 | ** |
| mean fold. depth | 7.7 | *** | 5.9 |  | 6.1 | ** |
| waist depth | 7 | *** | 5.2 |  | 5.7 | *** |
| % switches | 17 | * | 15 | *** | 11 | *** |
| unfiled files | 3 | *** | 8.3 |  | 5.9 | ** |
| files per folder | 6 | *** | 9 | * | 11 | *** |
| % empty folders | 8 | ** | 13 | *** | 5 | *** |
| mean file depth | 7 | *** | 5.5 |  | 5.8 | *** |
| file waist depth | 6.5 | *** | 5.2 |  | 5.2 | ** |

**Table 2:** Average measures of the scale and structure of file collections managed with different operating systems. Significance of pair-wise differences is marked between columns (the right-most column refers to Mac and Linux differences). Significance key: *** $p<0.0001$, ** $p<0.001$, * $p<0.05$

Despite Mac and Linux collections being comparable in storage space used (Windows collections are smaller), Mac collections have roughly double the files and three to four times the folders of Windows and Linux collections (i.e., a much larger overall scale). Though all tree structures display a wide, diamond shape [9] and have similarly deep waists (i.e., broadest parts) and mean folder depths, Mac collections' greater size is reflected slightly in maximum depth (17 vs 14) and greatly in breadth (5k vs 1k-1.4k). Windows collections have more root folders (22 as compared with 15-16), suggesting a slightly wider top to the diamond, and accordingly they have a slightly lower branching factor than Mac collections (i.e., 3.8 vs 5 subfolders per folder).

As a result of the larger folder tree, despite having more files, Mac collections have the fewest files per folders (6), roughly half that of Linux collections (11) or two-thirds of Windows collections (9). Linux collections feature significantly fewer switch folders than Mac or Windows collections (i.e., 11% vs 15-17% of folders contained only other folders and are thus presumed to be used only for navigation), and significantly fewer empty folders (5% vs 8% for Mac and 13% for Windows). While all OS's collections have relatively few unfiled files (i.e., files at the root), Mac collections featured the fewest (3 vs 6-8).

All but four Mac users and four Windows users (i.e., >99% of each group) reported using the OS-default file managers, while the 44 Linux participants used 14 different file managers; hence, meaningful analysis of differences across file managers was impossible.

## Discussion

RQ1 asked if the Big 5 personality traits were associated with or predictive of measures of file collections' scale or structure, and RQ2 asked the same question for sense of direction. We found only very weak associations and predictive models. This complements past work [2] that found no notable and significant associations of personality traits with file *retrieval*, and because work looking at personality traits and collections [20] also found only a few associations and in constrained conditions, we take these findings all together to suggest personality is not a notable (singular) determinant of FM. This implication is somewhat counter-intuitive given that many past studies conclude that PIM and FM practices are highly personal, and some associations have been observed between personality style and information behaviour more generally [15, 16]. PIM is personal, but it is also highly circumstantial, so examining variables collectively (e.g., for latent profiles) may prove more elucidating.

That sense of direction does not have notable associations with file collections' properties is surprising, especially given the related biological and cognitive supports discussed in prior works [1, 22] as outlined above. It is possible that command-line navigation is more demanding of spatial cognition than current graphical file navigation tools, for example because the tree view of folders aids navigation like a map would [23].

RQ3 sought to identify differences between collections managed with differing operating systems or file managers. As the vast majority of Mac and Windows users use the default file managers, we are not able to distinguish any fine-grained effects of FM software from the general effects of the OS, among which there are considerable differences. Like [3], we found Mac collections have, per folder, fewer files and more subfolders, but we did not observe those collections keeping more files or folders near the root of the tree; rather, Mac collections are much larger (in files) and perhaps as a consequence are slightly deeper (on average and at the tree bottoms) and much broader. The cause of these differences is currently unclear, as we are not aware of a clear difference between the operating systems or file managers that would have such an effect, and neither differences in hardware nor demographics provide further insight (e.g., Linux users were not all developers, which would have explained some of the differences [9, 7]). Considering the large effect sizes, further investigation is warranted, and to start we suggest analysing the different views and features provided by the file managers [3].

In summary, given the large differences in file collections across operating system reported here, and the lack of associations between collections and individual differences, we suggest future work explore potential technological determinants. It may also be fruitful to search FM collection and retrieval data (e.g., with clustering or machine learning techniques) for latent common profiles [21] to aid in understanding and modelling FM activity.

## Limitations

Our data do not reflect collection change over time nor users' perceptions (discussed further in [8]). As users' perceptions of organisation are connected to the number and depth of their folders [5], a triangulation of user perceptions and collection metrics would be useful.

## Conclusion

Supporting computer users in the daily task of managing files requires understanding what influences users' behaviour and constrains their task. The results presented here suggest personality and spatial ability are less useful than expected for improving that understanding, and that the tools utilised play a larger role than previously expected. Notably, Mac users seem to have much larger file collections than both Windows and Linux users, resulting in deeper and much broader folder trees. As a result, we recommend that studies seeking to understand and model FM identify common profiles in FM collection data and explore factors alternative to the individual differences explored here, such as the tools used in FM, external tasks and constraints (e.g., limited time or organisational policy), and the effects of these on the user experience.

## Acknowledgements
The authors thank the many participants for their time and effort, and Charles-Antoine Julien for his feedback.